\documentclass[conference]{IEEEtran}

\usepackage{amsfonts}
\usepackage{epsfig} 
\usepackage{amsmath} 
\usepackage{graphicx}

\newtheorem{theorem}{Theorem}[section]
\newtheorem{lemma}[theorem]{Lemma}
\newtheorem{definition}[theorem]{Definition}

\newtheorem{assumption}[theorem]{Assumption}

\newcommand{\sr}{\stackrel}

\newcommand{\tri}{\sr{\triangle}{=}}

\newcommand{\noi}{\noindent}

\newcommand{\be}{\begin{equation}}
\newcommand{\ee}{\end{equation}}
\newcommand{\bea}{\begin{eqnarray}}
\newcommand{\eea}{\end{eqnarray}}
\newcommand{\bes}{\begin{eqnarray*}}
\newcommand{\ees}{\end{eqnarray*}}

\newcommand{\bfi}{\begin{figure}}
\newcommand{\bfit}{\begin{figure}[t]}
\newcommand{\bfib}{\begin{figure}[b]}
\newcommand{\bfih}{\begin{figure}[h]}
\newcommand{\bfip}{\begin{figure}[p]}
\newcommand{\efi}{\end{figure}}

\newcommand{\bi}{\begin{itemize}}
\newcommand{\ei}{\end{itemize}}
\newcommand{\ben}{\begin{enumerate}}
\newcommand{\een}{\end{enumerate}}

\ifCLASSINFOpdf
\else
\fi

\ifCLASSINFOpdf
\else
\fi

\begin{document}
%
\title{Causal Rate Distortion Function on Abstract Alphabets: Optimal Reconstruction and Properties}

\author{\IEEEauthorblockN{\bf Photios A. Stavrou, Charalambos D. Charalambous and Christos K. Kourtellaris}
\IEEEauthorblockA{ECE Department, University of Cyprus, Green Park, Aglantzias 91,\\
 P.O. Box 20537, 1687, Nicosia, Cyprus\\
e-mail:\it~stavrou.fotios@ucy.ac.cy, chadcha@ucy.ac.cy, kourtellaris.christos@ucy.ac.cy}}


\maketitle

\begin{abstract}
A causal rate distortion function with a general fidelity criterion is formulated on abstract alphabets and a coding theorem is derived. Existence of the minimizing kernel is shown using the topology of weak convergence of probability measures. The optimal reconstruction kernel is derived, which is causal, and certain properties of the causal rate distortion function are presented.
\end{abstract}


\IEEEpeerreviewmaketitle

\section{INTRODUCTION}

\par Given a distortion or fidelity constraint between source  sequences $X^{\infty}\tri\{X_i\}_{i=0}^{\infty}\in{\cal X}^{\infty}\tri\times_{i=0}^{\infty}{\cal X}_i$ and reproduction sequences $Y^{\infty}\tri\{Y_i\}_{i=0}^{\infty}\in{\cal Y}^{\infty}\tri\times_{i=0}^{\infty}{\cal Y}_i$, non-causal codes achieve the rate distortion function (RDF) of the source, which is the optimal performance theoretically attainable. The RDF is described in \cite{thomas-cover91} for memoryless sources, in \cite{gallager} for stationary ergodic sources, in \cite{ihara1993} for information and distortion stable processes, and in \cite{han93} using the information spectrum method. The RDF for general sources on Polish spaces (complete separable metric spaces) and its properties are discussed extensively in \cite{csiszar74}.
\par Causal codes as defined in \cite{neuhoff1982} are a sub-class of non-causal codes, with the addition constraint on the reproduction coder (cascade of encoder-decoder) such that $Y_i$ depends on the past and present source symbols $\{X_0,X_1,\ldots,X_i\}$ but not on the future symbols $\{X_{i+1},X_{i+2},\ldots\}$, thus, $Y_i=f_i(X_0,X_1,\ldots,X_i)$ $\forall{i}$, where $\{f_i\}_{i=0}^{\infty}$ are measurable functions called reproduction coders.\\ 
Causal codes are extensively analyzed in \cite{neuhoff1982} using entropy type criteria (entropy of reproduction coder), further investigated in \cite{weissman2003} where side information is present, while \cite{linder2006} consider stationary sources at high resolution. The rate loss due to causality for Gaussian stationary sources with memory and mean square distortion is analyzed in \cite{gorbunov91}, and recently in \cite{derpich2011}.\\
Zero-delay codes are a sub-class of causal codes, with the additional constraint on the reproduction coder such that the reproduction $Y_i$ is done at the same time the corresponding source symbol $X_i$ is encoded, that is, both encoding and decoding are done causally. Sequential codes as defined in \cite{tatikonda00} and applied in \cite{tatikonda2004} are causal zero-delay codes such that the reproduction of each source symbol is done sequentially following the time ordering $X_0,Y_0,X_1,Y_1,\ldots$.\\
The objective of this paper is to impose a causal constraint on the reproduction coder, and formulate the causal source coding problem with fidelity criterion via rate distortion theory, on general alphabets using the topology of weak convergence of probability measures. The results include the following.
\begin{itemize}
\item[1)] Information theoretic definition of causal rate distortion function as an optimization problem in which the reproduction conditional distribution satisfies a causality constraint.
\item[2)] Source coding theorem for directed information and distortion stable processes.
\item[3)] Expression of the optimal causal reconstruction distribution and properties of the causal RDF.
\end{itemize}
\noi{\bf Causal Rate Distortion Function (CRDF)}.\\
The precise definition of causal codes is stated below and it is found in \cite{neuhoff1982}.
\begin{definition}(Causal Reproduction Coder)\label{causal_coder}
A reproduction coder is called causal if for all $i\leq{n}$
\begin{align}
f_i(x^n)=f_i(\tilde{x}^n)~\mbox{whenever}~x^i=\tilde{x}^i\nonumber
\end{align}
A source code is called causal if its induced reproduction coder is causal.
\end{definition}
From Definition~\ref{causal_coder}, it follows that the reproduction coder is causal if and only if the following Markov chain holds $(X_{i+1},X_{i+2},\ldots)\Leftrightarrow(X^i,Y^{i-1})\Leftrightarrow{Y_i}$, $i=0,1,\ldots$.\\
Assume an average distortion constraint
\begin{align}
E\big\{d_{0,n}(X^n,Y^n)\big\}\leq{D},\:d_{0,n}(x^n,y^n)\tri\frac{1}{n+1}\sum^n_{i=0}\rho_{i}(x^i,y^i)\nonumber 
\end{align}
where $D\geq0$,  $d_{0,n}(\cdot,\cdot)$  a non-negative distortion function.\\
Consider causal reproduction coders defined in Definition~\ref{causal_coder}. Define the causal convolution of conditional distributions by
\begin{align}
{\overrightarrow P}_{Y^n|X^n}(dy^n|x^n)\tri\otimes^n_{i=0}P_{Y_i|Y^{i-1},X^i}(dy_i|y^{i-1},x^i)\nonumber
\end{align}
Since a reproduction coder is causal if and only if the above Markov chain holds, then the reproduction conditional distribution of a causal coder satisfies
\begin{align}
{P}_{Y^n|X^n}(dy^n|x^n)={\overrightarrow P}_{Y^n|X^n}(dy^n|x^n),~P-a.s\label{exis5}
\end{align}
Substituting (\ref{exis5}) into mutual information $I(X^n;Y^n)$ it follows that for causal coders the information theoretic RDF for which an operational meaning will be saught, is given by
\begin{align}
R^c(D)&=\lim_{n\rightarrow\infty}\inf_{{\overrightarrow P}_{Y^n|X^n}:E\big\{d_{0,n}(X^n,Y^n)\big\}\leq{D}}\frac{1}{n+1}\int_{{\cal X}_{0,n}\times{\cal Y}_{0,n}}\nonumber\\
&\log\Big(\frac{{\overrightarrow P}_{Y^n|X^n}(dy^n|x^n)}{{P}_{Y^n}(dy^n)}\Big){\overrightarrow P}_{Y^n|X^n}(dy^n|x^n)  P_{X^n}(dx^n) \nonumber\\
& =\lim_{n\rightarrow\infty}\inf\frac{1}{n+1}{\mathbb I}_{X^n{\rightarrow}Y^n}(P_{X^n},{\overrightarrow P}_{Y^n|X^n})\label{exis9}
\end{align}
where the joint distribution $P_{X^n,Y^n}(dx^n,dy^n)$ for causal codes is uniquely defined by $P_{X^n,Y^n}(dx^n,dy^n)={\overrightarrow P}_{Y^n|X^n}(dy^n|x^n)\otimes{P}_{X^n}(dx^n)$. Note that (\ref{exis9}) is precisely the expression consider in \cite{tatikonda00} to derive coding theorem for sequential codes. It is easy to verify that ${\mathbb I}_{X^n{\rightarrow}Y^n}(P_{X^n},{\overrightarrow P}_{Y^n|X^n})$ is the directed information from $X^n$ to $Y^n$, $I(X^n\rightarrow{Y^n})\tri\sum_{i=0}^n{I}(X^i;Y_i|Y^{i-1})$, subject to the requirement that the source is not affected by past reconstruction symbols, that is, $\otimes_{i=0}^n{P}_{X_i|X^{i-1},Y^{i-1}}(dx_i|x^{i-1},y^{i-1})=\otimes_{i=0}^n{P}_{X_i|X^{i-1}}(dx_i|x^{i-1})=P_{X^n}(dx^n)$. However, if the causality constraint (\ref{exis5}) is not imposed, the conditional distribution ${\overrightarrow P}_{Y^n|X^n}(dy^n|x^n)$ in (\ref{exis9}) should be replaced by ${P}_{Y^n|X^n}(dy^n|x^n)$, and the resulting expression is the classical RDF. Since, by the chain rule  ${P}_{Y^n|X^n}(dy^n|X^n=x^n)=\otimes^n_{i=0}P_{Y_i|Y^{i-1}=y^{i-1},X^n=x^n}(dy_i|Y^{i-1}=y^{i-1},X^n=x^n)$, in general the classical RDF solution yields reconstructions of $Y_i=y_i$ which depends on future values of the source symbols $(X_{i+1}=x_{i+1},\ldots,X_n=x_n)$, in addition to its past reconstruction symbols $Y^{i-1}=y^{i-1}$, and past and present symbols $X^i=x^i$. On the other hand (\ref{exis5}) implies causality.
\section{PROBLEM FORMULATION AND CODING THEOREMS}
Let $\mathbb{N}^n\tri\{0,1,\ldots,n\}$, $n \in \mathbb{N} \tri \{0,1,2,\ldots\}$. The source and reconstruction alphabets are sequences of Polish spaces \cite{dupuis-ellis97} $\{ {\cal X}_t: t\in\mathbb{N}\}$ and $\{ {\cal Y}_t: t\in\mathbb{N}\}$, respectively, (e.g., ${\cal Y}_t, {\cal X}_t$ are complete separable metric spaces), associated with their corresponding measurable spaces $({\cal X}_t,{\cal B}({\cal X}_t))$ and $({\cal Y}_t, {\cal B}({\cal Y}_t))$. Sequences of alphabets are  identified with the product spaces $({\cal X}_{0,n},{\cal B}({\cal X}_{0,n})) \tri  \times_{k=0}^{n}({\cal X}_k,{\cal B}({\cal X}_k))$,
and $({\cal Y}_{0,n},{\cal B}({\cal Y}_{0,n}))\tri \times_{k=0}^{n}({\cal Y}_k,{\cal B}({\cal Y}_k))$.
The source and reconstruction are processes denoted by $X^n \tri \{X_n: n\in\mathbb{N}^n\}$, $X_n\in{\cal X}_n$, and by $Y^n \tri \{Y_n: n\in\mathbb{N}^n\}$, $Y_n\in{\cal Y}_n$, respectively. Probability measures on any measurable space  $( {\cal Z}, {\cal B}({\cal Z}))$ are denoted by ${\cal M}_1({\cal Z})$. It is assumed  that the $\sigma$-algebras $\sigma\{X^{-1}\}=\sigma\{Y^{-1}\}=\{\emptyset,\Omega\}$.
\begin{definition}\label{stochastic kernel}
Let $({\cal X}, {\cal B}({\cal X})), ({\cal Y}, {\cal B}({\cal Y}))$ be measurable spaces in which $\cal Y$ is a Polish Space.\\
A  stochastic Kernel on $\cal Y$ given $\cal X$ is a mapping $q: {\cal B}({\cal Y}) \times {\cal X}  \rightarrow [0,1]$ satisfying the following two properties:
\par 1) For every $x \in {\cal X}$, the set function $q(\cdot;x)$ is a probability measure (possibly finitely additive) on ${\cal B}({\cal Y}).$
\par 2) For every $F \in {\cal B}({\cal Y})$, the function $q(F;\cdot)$ is ${\cal B}({\cal X})$-measurable.\\
 The set of all such stochastic Kernels is denoted by ${\cal Q}({\cal Y};{\cal X})$.
\end{definition}
\par Stochastic kernels are classified into non-causal and causal as follows.
\begin{definition}\label{comprchan}
Given measurable spaces $({\cal X}_{0,n},{\cal B}({\cal X}_{0,n}))$, $({\cal Y}_{0,n},{\cal B}({\cal Y}_{0,n}))$, and their product spaces, data compression channels are defined as follows.
\begin{enumerate}
\item {\it A Non-Causal Data Compression Channel} is a  stochastic kernel $ q_{0,n} (dy^n; x^n) \in {\cal Q}({\cal Y}_{0,n} ;{\cal X}_{0,n}), n \in \mathbb{N}$.
\item{\it A Causal Product Data Compression Channel} is a convolution of a sequence of causal stochastic kernels defined by
\bes
{\overrightarrow q}_{0,n}(dy^n;x^n)&\tri\otimes_{i=0}^n q_i(dy_i;y^{i-1},x^i)
\ees
\end{enumerate}
where $q_i \in {\cal Q}({\cal Y}_i;{\cal Y}_{0,i-1}\times{\cal X}_{0,i}), i=0,\ldots,n,~n \in \mathbb{N}$. The set of such convolution of causal kernels is denoted by $\overrightarrow{\cal Q}({\cal Y}_{0,n};{\cal X}_{0,n})$.
\end{definition}
\subsection{Information Theoretic Causal Rate Distortion Function}
This section gives the abstract formulation of $R^c(D)$. Given a source probability measure ${\cal \mu}_{0,n} \in {\cal M}_1({\cal X}_{0, n})$ and a reconstruction kernel ${\overrightarrow q}_{0,n} \in \overrightarrow{\cal Q}({\cal Y}_{0, n};{\cal X}_{0, n})$ consistent with causal reproduction coder, define the following probability measures.\\
{\bf P1}: The joint measure $P_{0,n} \in {\cal M}_1({\cal Y}_{0,n}\times {\cal X}_{0, n})$:
\begin{align}
P_{0,n}(G_{0,n})&\tri(\mu_{0,n} \otimes {\overrightarrow q}_{0,n})(G_{0,n}),\:G_{0,n} \in {\cal B}({\cal X}_{0,n})\times{\cal B}({\cal Y}_{0,n})\nonumber\\
&=\int_{{\cal X}_{0,n}} {\overrightarrow q}_{0,n}(G_{0,n,x^n};x^n) \mu_{0,n}(d{x^n})\nonumber
\end{align}
where $G_{0,n,x^n}$ is the $x^n-$section of $G_{0,n}$ at point ${x^n}$ defined by $G_{0,n,x^n}\tri \{y^n \in {\cal Y}_{0,n}: (x^n, y^n) \in G_{0,n}\}$ and $\otimes$ denotes the convolution.\\
{\bf P2}: The marginal measure $\nu_{0,n} \in {\cal M}_1({\cal Y}_{0,n})$:
\begin{align}
\nu_{0,n}(F_{0,n})&\tri P_{0,n}({\cal X}_{0, n} \times F_{0,n}),~F_{0,n} \in {\cal B}({\cal Y}_{0,n})\nonumber\\
&=\int_{{\cal X}_{0, n}} {\overrightarrow q}_{0,n}(F_{0,n};x^n) \mu_{0,n}(dx^n)\nonumber
\end{align}
{\bf P3}: The product measure  $\pi_{0,n}:{\cal B}({\cal X}_{0,n}) \times
{\cal B}({\cal Y}_{0,n}) \mapsto [0,1] $ of $\mu_{0,n}\in{\cal M}_1({\cal X}_{0, n})$ and $\nu_{0,n}\in{\cal M}_1({\cal Y}_{0, n})$:
\begin{align}
\pi_{0,n}(G_{0,n})&\tri(\mu_{0,n} \times \nu_{0,n})(G_{0,n}),~G_{0,n} \in {\cal B}({\cal X}_{0,n}) \times {\cal B}({\cal Y}_{0,n})\nonumber\\
&=\int_{{\cal X}_{0, n}} \nu_{0,n}(G_{0,n,x^n}) \mu_{0,n}(dx^n)\nonumber
\end{align}
\noi The precise information measure used to define CRDF is the mutual information between two sequences of random processes $X^n$ and $Y^n$ whose distributions are consistent with the definition of the causal reproduction coder, e.g., generated via ${\bf P1}$-${\bf P3}$. Hence, by the construction of probability measures ${\bf P1}$-${\bf P3}$, and the chain rule of relative entropy \cite{dupuis-ellis97}:
\begin{align}
&I(X^n;Y^n) \tri  \mathbb{D}(P_{0,n}|| \pi_{0,n})\label{eq1}\\
&=\int_{{\cal X}_{0,n} \times {\cal Y}_{0,n}}\log \Big( \frac{d  (\mu_{0,n} \otimes {\overrightarrow q}_{0,n}) }{d ( \mu_{0,n} \times \nu_{0,n} ) }\Big) d(\mu_{0,n} \otimes {\overrightarrow q}_{0,n}) \nonumber\\
&= \int\log \Big( \frac{{\overrightarrow q}_{0,n}(d y^n; x^n)}{  \nu_{0,n} (dy^n)   } \Big){\overrightarrow q}_{0,n}(dy^n;x^n) \mu_{0,n}(dx^n) \label{exis6}\\
&\equiv \mathbb{I}_{X^n\rightarrow{Y^n}}(\mu_{0,n},{\overrightarrow q}_{0,n})  \label{re3}
\end{align}
Note that $(\ref{re3})$ states that mutual information is expressed as a functional of $\{\mu_{0,n}, {\overrightarrow q}_{0,n}\}$ denoted by $\mathbb{I}_{X^n\rightarrow{Y^n}}(\mu_{0,n},{\overrightarrow q}_{0,n})$. Also, if the causality assumption on the reproduction coder is not imposed, then $I(X^n;Y^n)=\mathbb{I}(\mu_{0,n},{q}_{0,n})$, which is how classical RDF is defined.
\par The next lemma gives equivalent statements which are consistent with causal reproduction coders in terms of causal convolution of reconstruction kernels, mutual information, directed information, and conditional independence.
\begin{lemma} \label{lem1}
The following are equivalent for each $n\in\mathbb{N}$.
\begin{enumerate}
\item $q_{0,n} (dy^n; x^n)={\overrightarrow q}_{0,n}(dy^n;x^n)$ a.s., where ${\overrightarrow q}_{0,n}$ is given in Definition \ref{comprchan}-2).

\item For each $i=0,1,\ldots, n-1$,  $Y_i \Leftrightarrow (X^i, Y^{i-1}) \Leftrightarrow (X_{i+1}, X_{i+2}, \ldots, X_n)$, forms a Markov chain.

\item $I(X^n ; Y^n)=I(X^n \rightarrow Y^n)$.


\item For each  $i=0,1,\ldots, n-1$, $Y^i \Leftrightarrow X^i \Leftrightarrow X_{i+1}$ forms a Markov chain.
\end{enumerate}
\end{lemma}
{\IEEEproof}The prove is omitted due to space limitation.{\endIEEEproof}
$I(X^n; Y^n)=I(X^n\rightarrow{Y}^n)\equiv{\mathbb{I}}_{X^n\rightarrow{Y^n}}(\mu_{0,n},\overrightarrow{q}_{0,n} )$ is a functional of $\{\mu_{0,n},{\overrightarrow q}_{0,n}\}$.
Hence, the information definition of a causal rate distortion is defined by optimizing ${\mathbb I}(\mu_{0,n},\overrightarrow{q}_{0,n})$ over ${\overrightarrow q}_{0,n}$ which satisfies a distortion constraint.
\begin{definition}\label{def1}
(Causal Information Rate Distortion Function)
Suppose $d_{0,n}(x^n,y^n)\tri\frac{1}{n+1}\sum^n_{i=0}\rho_{i}(x^i,y^i)$, where $\rho_{i}: {\cal X}_{i}  \times {\cal Y}_{i}\rightarrow [0, \infty)$, is a sequence of ${\cal B}({\cal X}_{i}) \times {\cal B }( {\cal Y}_{i})$-measurable distortion functions, and let $\overrightarrow{Q}_{0,n}(D)$ (assuming is non-empty) denotes the average distortion or fidelity constraint defined by
\begin{align}
\overrightarrow{Q}_{0,n}&(D)\tri\Big\{\overrightarrow{q}_{0,n} \in  \overrightarrow{\cal Q}({\cal Y}_{0,n};{\cal X}_{0,n}):\ell({\overrightarrow{q}}_{0,n})\tri\nonumber\\
&\frac{1}{n+1}\int_{{\cal X}_{0,n}\times{{\cal Y}}_{0,n}}d_{0,n}({x^n},{y^n})(\overrightarrow{q}_{0,n}\otimes\mu_{0,n})(d{x}^{n},d{y}^{n})\nonumber\\
&\leq D\Big\},~D\geq0 \label{eq2}
\end{align}
Define
\bea
{R}_{0,n}^c(D) \tri  \inf_{{\overrightarrow{q}_{0,n}\in \overrightarrow{Q}_{0,n}(D)}}\frac{1}{n+1}\mathbb{I}_{X^n\rightarrow{Y^n}}(\mu_{0,n},{\overrightarrow q}_{0,n})
\label{ex12}
\eea
The operational meaning of CRDF is established via ${R}^c(D)\tri\lim_{n\rightarrow{\infty}}{R}_{0,n}^c(D)$, provided the limit exists.
\end{definition}
Clearly, ${R}_{0,n}^c(D)$ is characterized by minimizing $\mathbb{I}_{X^n\rightarrow{Y^n}}(\mu_{0,n},{\overrightarrow q}_{0,n})$ over the causal convolution measure ${\overrightarrow q}_{0,n}\in{\overrightarrow Q}_{0,n}(D)$.
\subsection{Coding Theorems for Causal and Sequential Codes}
This section gives an operational meaning to ${R}^c_{0,n}(D)$ via coding theorems. There are two cases, sequential codes and causal codes.\\
{\bf Sequential Codes}. Coding theorems for sequential codes are established in \cite{tatikonda00} for the finite alphabet case, and two-dimensional source $X^{T,N}\tri\{X_{t,n}:t=0,\ldots,T,n=0,\ldots,N\}$, where $t$ represents time index and $n$ represents spatial index, under the assumption that $P(X^{T,N})=\otimes_{n=0}^N{P}(X_n^T)$, and $\{X_n^T:n=0,\ldots,N\}$ are identically distributed, and the distortion constraint is $E_{X^{T,N}}\big\{\frac{1}{N+1}\sum_{n=0}^N\rho(x_{t,n},y_{t,n})\leq{D}_t,~t=0,1,\ldots,T\big\}$. With a slight modification of the per-letter distortion function above, it can be shown  that the coding theorem in  \cite{tatikonda2004} is still valid, and that the corresponding sequential RDF is
given by $R^{SRD}(D)\equiv{R}^c(D)$. The coding theorem is derived using strong typicality.\\
{\bf Causal Codes}. Here we describe a coding theorem for causal codes.
\begin{definition}(Causal Code)\label{causal_code}
A $(n,2^{nR},D)$ causal source code of block length $n$, and rate R consists of an encoding mapping $e(\cdot)$,
$e:{\cal X}_{0,n}\longrightarrow{\cal W}\tri\{1,2,\ldots,2^{nR}\}$
and a sequence of decoder mapping $\{g_i\}_{i=0}^n(\cdot)$,
$g_i:\{1,2,\ldots,2^{nR}\}\longrightarrow{\cal Y}_i,~i=0,1,\ldots,n$
such that the sequence of reproduction coders $\{f_i=g_i\circ{e}\}_{i=0}^n$ are causal.
\end{definition}
\begin{definition}(Achievable Rate)\label{achievable_rate}
A rate distortion pair $(R,D)$ is called achievable if $\forall\epsilon>0$ and sufficiently large $n$ there exists a $(n,2^{nR},D)$ causal code such that 
\begin{align*}
\frac{1}{n+1}E\big\{d_{0,n}(X^n,Y^n)\big\}\leq{D}+\epsilon
\end{align*}
\end{definition}
\begin{definition}(Causal Rate Distortion Function)\label{CRDF}
The CRDF $R(D)$ is the infimum of rates $R$ such that $(R,D)$ is achievable.
\end{definition}
The definition of the coding theorem can be done for i) stationary ergodic processes $\big\{(X_i,Y_i):i=0,1,\ldots\big\}$ by invoking versions of Shannon-McMillan-Breimann Theorem,
ii) for information and distortion stable processes by invoking versions of Dobrushin's conditions, and iii) for processes with information spectrum via variants of the methods in \cite{han93}.\\
Here we discuss ii) since the distortion function $d_{0,n}(x^n,y^n)$ is general and does not fall under the special case discussed in \big{[}\cite{gallager}, Section 9.8\big{]} for ergodic sources.\\
Define the information density consistent with the causal reproduction coder by ${\Lambda}_{0,n}(x^n,y^n)\tri\log\frac{{\overrightarrow{P}}_{Y^n|X^n}(dy^n|x^n)}{P_{Y^n}(dy^n)}$
where it is assumed absolute continuity ${\overrightarrow{P}}_{Y^n|X^n}(\cdot|x^n)\ll{P}_{Y^n}(\cdot)$, $\mu_{0,n}-a.s.$ for almost all $x^n\in{\cal X}_{0,n}$. Then $\mathbb{I}_{X^n\rightarrow{Y^n}}(P_{X^n},{\overrightarrow P}_{Y^n|X^n})=E\big\{\Lambda_{0,n}(x^n,y^n)\big\}$, where the joint distribution is $P_{X^n,Y^n}=P_{X^n}\otimes{\overrightarrow P}_{Y^n|X^n}$.
\begin{definition}(Information and Distortion Stable)\label{stable}
For each $\epsilon>0$ define the $\epsilon$-typical set of directed information density
\begin{align*}
{\cal T}_{\epsilon}^{(n)}&\tri\bigg\{(x^n,y^n)\in{\cal X}_{0,n}\times{\cal Y}_{0,n}:\bigg{|}\frac{1}{n+1}\log\frac{{\overrightarrow{P}}_{Y^n|X^n}(dy^n|x^n)}{P_{Y^n}(dy^n)}\\
&-\frac{1}{n+1}\mathbb{I}_{X^n\rightarrow{Y^n}}(P_{X^n},{\overrightarrow P}_{Y^n|X^n})\bigg{|}<\epsilon\bigg\}
\end{align*}
and the $\epsilon$-typical set of the distortion by
\begin{align*}
{\cal D}_{\epsilon}^{(n)}&\tri\bigg\{(x^n,y^n)\in{\cal X}_{0,n}\times{\cal Y}_{0,n}:\bigg{|}\frac{1}{n+1}d_{0,n}(x^n,y^n)\\
&-\frac{1}{n+1}E\big\{d_{0,n}(x^n,y^n)\big\}\bigg{|}<\epsilon\bigg\}
\end{align*}
\end{definition}
The process $\big\{(X_n,Y_n):n\in\mathbb{N}\big\}$ is called directed information and distortion stable if $
\lim_{n\rightarrow\infty}Prob({\cal T}_{\epsilon}^{(n)})=1,$ and $\lim_{n\rightarrow\infty}Prob({\cal D}_{\epsilon}^{(n)})=1$, respectively, for every $\epsilon>0$.\\
Note that for stationary ergodic process $\big\{(X_n,Y_n):n\in\mathbb{N}\big\}$ and certain distortion functions (see \cite{gallager}, Section 9.8) information and distortion stability follows.
Before the statements leading to coding theorem are introduced, the notion of stability of the source is required.
\begin{definition}\label{source_stable}
The source $\{X_n:n\in\mathbb{N}\}$ is called stable if for any given $D>0$ and $\epsilon>0$ there exists $\{Y_n:n\in\mathbb{N}\}$ such that $\big\{(X_n,Y_n):n\in\mathbb{N}\big\}$ is directed information and distortion stable, and
\begin{align}
&\lim_{n\rightarrow\infty}\frac{1}{n+1}E\big\{d_{0,n}(x^n,y^n)\big\}\leq{D}\label{ex7}\\
&\lim_{n\rightarrow\infty}\frac{1}{n+1}E\Big\{\log\frac{{\overrightarrow{P}}_{Y^n|X^n}(dy^n|x^n)}{P_{Y^n}(dy^n)}\Big\}\leq{R^c(D)}+\epsilon\label{ex8}
\end{align} 
where $E\{\cdot\}$ is with respect to $P_{X^n,Y^n}=\overrightarrow{P}_{Y^n|X^n}\otimes{P}_{X^n}$.
\end{definition}
Note that by specializing $d_{0,n}(x^n,y^n)$ to distortion functions that satisfy sub-additivity property the limit in (\ref{ex7}) exists.
Utilizing Definition~\ref{source_stable}, it can be shown that the following statements hold, which are vital in establishing the coding theorem.
\begin{lemma}\label{helpful_lemma}
Assume $\{X_n:n\in\mathbb{N}\}$ is stable and the joint distribution $P_{X^n,Y^n}$ is defined by $P_{X^n,Y^n}(dx^n,dy^n)=\overrightarrow{P}_{Y^n|X^n}(dy^n|x^n)\otimes{P}_{X^n}(dx^n)$.\\
Then
\begin{itemize}
\item[1)]$\lim_{n\rightarrow\infty}P_{X^n,Y^n}({\cal T}_{\epsilon}^{(n)})=\lim_{n\rightarrow\infty}P_{X^n,Y^n}({\cal D}_{\epsilon}^{(n)})=1$
\item[2)]For sufficiently large $n$, there exists $\epsilon>0$ such that 
\begin{align*}
\frac{\overrightarrow{P}_{Y^n|X^n}(dy^n|x^n)}{{P}_{X^n}(dx^n)}\leq{2}^{n\big(\mathbb{I}_{X^n\rightarrow{Y^n}}({P}_{X^n},\overrightarrow{P}_{Y^n|X^n})+3\epsilon\big)}
\end{align*}
\end{itemize}
\end{lemma}
Using Lemma~\ref{helpful_lemma}, the source coding theorem stated below can be established.
\begin{theorem}(Source Coding Theorem)
Assume $\{X_n:n\in\mathbb{N}\}$ is stable and $\sup_{(x^i,y^i)\in{\cal X}_{0,i}\times{\cal Y}_{0,i}}\rho_i(x^i,y^i)<k$, $k<\infty$ for all $i$. If $R>{R}^c(D)$ then for any $\delta>0$ and sufficiently large $n$, there exists an $(n,2^{nR},D)$ causal code which satisfies the average distortion $\frac{1}{n+1}E_{P_{X^n,Y^n}}\big\{\frac{1}{n+1}d_{0,n}(x^n,y^n)\big\}\leq{D}+\delta$.
{\IEEEproof} The derivation utilizes Lemma~\ref{helpful_lemma}, and random codebook generation. Fix $\overrightarrow{P}_{Y^n|X^n}(dy^n|x^n)$, which achieves the equality in $R^c(D)$ $\big{(}$e.g., (\ref{ex12})$\big{)}$. Calculate ${\overrightarrow P}_{Y^n}(dy^n)=\int_{{\cal X}_{0,n}}\overrightarrow{P}_{Y^n|X^n}(dy^n|x^n){P}_{X^n}(dx^n)$. Randomly generate rate distortion codebook ${\cal C}$ of $2^{nR}$ sequences $Y^n$ according to ${\overrightarrow P}_{Y^n}(dy^n)$ and reveal the codebook to encoder and decoder. Utilizing Lemma~\ref{helpful_lemma} and Definition~\ref{source_stable}, the result is obtained following \cite{ihara1993}.{\endIEEEproof}
\end{theorem}
\section{EXISTENCE OF OPTIMAL CAUSAL RECONSTRUCTION}
\par In this section, the existence of the minimizing causal product kernel in $(\ref{ex12})$ is shown by using the topology of weak convergence of probability measures on Polish spaces. The only assumptions required are 1) ${\cal Y}_{0,n}$ is a compact Polish space, 2) ${\cal X}_{0,n}$ is a Polish space, and  3) $d_{0,n}(x^n,\cdot)$ is continuous on ${\cal Y}_{0,n}$.
\subsection{Weak Compactness and Existence of Optimal Reconstruction Kernel}
\par Define the family of measures
\begin{align*}
{\overrightarrow{\cal Q}}({\cal Y}_{0,n};{\cal X}_{0,n})&=\big\{{\overrightarrow{q}}_{0,n}(dy^n;x^n):{\overrightarrow{q}}_{0,n}(dy^n;x^n)\nonumber\\
&=\otimes_{i=0}^n{q}_i(dy_i;y^{i-1},x^i)\big\}\nonumber
\end{align*}
\begin{lemma}\label{weak_compactness}
Let ${\cal Y}_{0,n}$ be a compact Polish space and ${\cal X}_{0,n}$ a Polish space.\\
Then
\begin{itemize}
\item[1)] The family of measures ${\overrightarrow{q}}_{0,n}(dy^n;x^n)\in{\overrightarrow{\cal Q}}({\cal Y}_{0,n};{\cal X}_{0,n})$ is compact.
\item[2)] Under the assumption that $d_{0,n}(x^n,\cdot)$ is continuous in ${\cal Y}_{0,n}$ the set ${\overrightarrow{Q}}_{0,n}(D)$ is a closed subset of ${\overrightarrow{\cal Q}}({\cal Y}_{0,n};{\cal X}_{0,n})$.
\end{itemize}
\end{lemma}
{\IEEEproof} 1) This follows from the fact that any ${\overrightarrow{q}}_{0,n}(dy^n;x^n)\in{\overrightarrow{\cal Q}}({\cal Y}_{0,n};{\cal X}_{0,n})$ is factorized as ${\overrightarrow{q}}_{0,n}(dy^n;x^n)=\otimes_{i=0}^n{q}_i(dy_i;y^{i-1},x^i)$, where $q_i(dy_i;y^{i-1},x^i)\in{\cal Q}({\cal Y}_i;{\cal Y}_{0,i-1}\times{\cal X}_{0,i})$, $1\leq{i}\leq{n}$, and ${\cal Y}_{0,n}$ compact Polish space implies that $\{q_i(\cdot;y^{i-1},x^i):y^{i-1}\in{\cal Y}_{0,i-1},x^i\in{\cal X}_{0,i}\}$ is compact, $\forall{i}$. Utilizing this, by induction it can be shown that the family of convolution measures ${\overrightarrow{\cal Q}}({\cal Y}_{0,n};{\cal X}_{0,n})$ is compact.\\
2) Utilizing compactness of ${\overrightarrow{\cal Q}}({\cal Y}_{0,n};{\cal X}_{0,n})$ and the assumption on  $d_{0,n}(x^n,\cdot)$ it can be shown that ${\overrightarrow{Q}}_{0,n}(D)$ is a closed subset of ${\overrightarrow{\cal Q}}({\cal Y}_{0,n};{\cal X}_{0,n})$. {\endIEEEproof}
\par The next theorem establishes existence of the minimizing reconstruction kernel for (\ref{ex12}). 
\begin{theorem}\label{existence_rd}
Suppose ${\cal Y}_{0,n}$ is compact Polish space and $d_{0,n}(x^n,\cdot)$ is continuous in ${\cal Y}_{0,n}$. Then ${R}^c_{0,n}(D)$ has a minimum.
\end{theorem}
{\IEEEproof} The assumptions are sufficient to show lower semicontinuity of the functional $\mathbb{I}_{X^n\rightarrow{Y^n}}(\mu_{0,n},{\overrightarrow q}_{0,n})$ with respect to ${\overrightarrow{q}}_{0,n}$ for a fixed ${\mu}_{0,n}$. Moreover, by Lemma~\ref{weak_compactness}, 2) since $\overrightarrow{Q}_{0,n}(D)$ is a closed subset of a compact set $\overrightarrow{\cal Q}({\cal Y}_{0,n};{\cal X}_{0,n})$, then $\overrightarrow{Q}_{0,n}(D)$ is also compact. By Weiestrass theorem existence follows.
{\endIEEEproof}
\section{OPTIMAL CAUSAL RECONSTRUCTION}
\par In this section the form of the optimal causal reconstruction kernel is derived and the properties of $R_{0,n}^c(D)$ are discussed under a stationarity assumption.
\subsection{Optimal Reconstruction}
\begin{assumption}\label{stationarity}
The family of measures that admits the factorization ${\overrightarrow q}(dy^n|x^n)=\otimes^n_{i=0}q_i(dy_i|y^{i-1},x^i)$ is the convolution of stationary conditional distributions.
\end{assumption}
Assumption~\ref{stationarity} holds for stationary ergodic process $\{(X_i,Y_i):i\in\mathbb{N}\}$ and $\rho_i(x^i,y^i)$, which is stationary and time-invariant $\forall{i}$. The method is based on calculus of variations on the space of measures \cite{dluenberger69}. Utilizing Assumption~\ref{stationarity}, which holds for stationary ergodic processes $\{(X_i,Y_i):i=0,1,\ldots,n\}$ and single letter distortion function or distortion function discussed in \big{[}\cite{gallager}, Section 9.8\big{]}, the Gateaux differential of $\mathbb{I}_{X^n\rightarrow{Y^n}}(\mu_{0,n},{\overrightarrow q}_{0,n})$ is done in only one direction $\big{(}$since $q_i(dy_i;y^{i-1},x^i)$ are stationary$\big{)}$. This simplifies the calculations of Gateaux derivative of $\mathbb{I}_{X^n\rightarrow{Y^n}}(\mu_{0,n},{\overrightarrow q}_{0,n})$.
\begin{theorem} \label{th5}
Suppose ${\mathbb I}_{\mu_{0,n}}({\overrightarrow q}_{0,n}) \tri \mathbb{I}_{X^n\rightarrow{Y^n}}(\mu_{0,n},{\overrightarrow q}_{0,n})$ is well defined for every ${\overrightarrow q}_{0,n}\in\overrightarrow{Q}_{0,n}(D)$ possibly taking values from the set $[0,\infty).$ Then  ${\overrightarrow q}_{0,n} \rightarrow {\mathbb I}_{\mu_{0,n}}({\overrightarrow q}_{0,n})$ is
Gateaux differentiable at every point in $\overrightarrow{Q}_{0,n}(D)$, and the Gateaux
derivative at the  point ${\overrightarrow q}_{0,n}^0$ in the direction ${\overrightarrow q}_{0,n}-{\overrightarrow q}_{0,n}^0$ is given
by
\begin{align}
&\delta{\mathbb I}_{\mu_{0,n}}({\overrightarrow q}_{0,n}^0;{\overrightarrow q}_{0,n}-{\overrightarrow q}_{0,n}^0)=\int_{{\cal X}_{0,n}\times{\cal Y}_{0,n}}\log \Big(
\frac{{\overrightarrow q}_{0,n}^0(dy^n;x^n)}{\nu_{0,n}^0(dy^n)}\Big)\nonumber\\
&({\overrightarrow q}_{0,n}-{\overrightarrow q}_{0,n}^0)(dy^n;x^n) \mu_{0,n}(dx^n)
\end{align}
where $\nu_{0,n}^0\in{\cal M}_1({\cal Y}_{0,n})$ is the marginal measure corresponding
to ${\overrightarrow q}_{0,n}^0\otimes\mu_{0,n}\in{\cal M}_1({\cal Y}_{0,n}\times{\cal X}_{0,n})$.
\end{theorem}
{\IEEEproof}The proof utilizes Assumption~\ref{stationarity}.{\endIEEEproof}
\par The constrained problem defined by (\ref{ex12}) can be reformulated using Lagrange multipliers as follows (equivalence of constrained and unconstrained problems follows from \cite{dluenberger69}).
\begin{align}
{R}_{0,n}^c(D) &= \sup_{s\leq0}\inf_{{\overrightarrow q}_{0,n} \in \overrightarrow{\cal Q}({\cal Y}_{0,n};{\cal X}_{0,n})}\Big\{ \frac{1}{n+1}\mathbb{I}_{X^n\rightarrow{Y^n}}(\mu_{0,n},{\overrightarrow q}_{0,n})\nonumber\\
&-s\big(\ell({\overrightarrow q}_{0,n})-D\big)\Big\} \label{ex13}
\end{align}
and  $s \in(-\infty,0]$ is the Lagrange multiplier.\\
Note that $\overrightarrow{\cal Q}({\cal Y}_{0,n};{\cal X}_{0,n})$ represents the causality constraint set. Therefore, one should introduce another set of Lagrange multipliers to obtain an optimization without constraints. This process is involved hence we state the main results.
\begin{theorem} \label{th6}
Suppose Assumption~\ref{stationarity} and $d_{0,n}(x^n,y^n)=\sum_{i=0}^n\rho_{i}(x^i,y^i)$ hold. The infimum in $(\ref{ex13})$ is attained at  $\overrightarrow{q}^*_{0,n} \in
\overrightarrow{Q}_{0,n}(D)$ given by
\begin{align}
\overrightarrow{q}^*_{0,n}(dy^n;x^n)=\otimes_{i=0}^n\frac{e^{s \rho_{i}(x^i,y^i)}
\nu^*_i(dy_i;y^{i-1})}{\int_{{\cal Y}_i} e^{s \rho_{i}(x^i,y^i)} \nu^*_i(dy_i;y^{i-1})}\label{ex14}
\end{align}
and $\nu^*_i(dy_i;y^{i-1})\in {\cal Q}({\cal Y}_i;{\cal Y}_{0,{i-1}})$. The causal rate distortion
function is given by
\begin{align}
&{R}_{0,n}^c(D)=sD - \frac{1}{n+1}\sum_{i=0}^n\int_{{{\cal X}_{0,i}}\times{{\cal Y}_{0,i-1}}}\log \Big(\nonumber\\
&\int_{{\cal Y}_i} e^{s\rho_{i}(x^i,y^i)} \nu^*_i(dy_i;y^{i-1})\Big){{\overrightarrow q}^*_{0,i-1}}(dy^{i-1};x^{i-1})\otimes \mu_{0,i}(dx^i)\label{ex15}
\end{align}
If ${R}_{0,n}^c(D) > 0$ then $ s < 0$  and
\bes
\frac{1}{n+1}\sum_{i=0}^n\int_{{\cal X}_{0,i}} \int_{{\cal Y}_{0,i}}
\rho_{i}(x^i,y^i) {\overrightarrow q}^*_{0,i}(dy^i;x^i) \mu_{0,i}(dx^i)=D
\ees
\end{theorem}
{\IEEEproof} The fully unconstraint problem of (\ref{ex13}) is obtained by introducing another set of Lagrange multipliers. Using this and Theorem~\ref{th5} we obtain (\ref{ex14}) and (\ref{ex15}).{\endIEEEproof}
Note that according to Assumption~\ref{stationarity}, the terms appear in the right side of (\ref{ex14}) are identical.
\subsection{PROPERTIES OF $R_{0,n}^c(D)$}
In this section, we present some important properties of the CRDF as it is defined in (\ref{ex12}).
\begin{theorem} \label{prop1}
\noi
\begin{enumerate}
\item ${R}_{0,n}^c(D)$ is a convex, non-increasing function of $D$
\item If $\rho_{i} \in L^1(\pi_{i})$ then \\
a) ${R}_{0,n}^c(\frac{1}{n+1}\sum_{i=0}^nE_{\pi_{i}}(\rho_{i}))=0$; \\
b) ${R}_{0,n}^c(D)$ is non-increasing for $D \in [0, D_{max}]$ where $D_{max}=\frac{1}{n+1}\sum_{i=0}^nE_{\pi_{i}}(\rho_{i})$ and ${R}_{0,n}^c(D)=0$ for any $D \geq D_{max}$
\item ${R}_{0,n}^c(D) > 0$ for all $D < D_{max}$ and ${R}_{0,n}^c(D)=0$ for all $D \geq D_{max}$, where
\bes
D_{max}= \min_{\{y^n\}\in {\cal Y}_{0,n}}\frac{1}{n+1}\sum_{i=0}^n \int_{{\cal X}_{0,i}} \rho_{i}(x^i,y^i)\mu_{0,i}(dx^i)
\ees
if such a minimum exists.
\end{enumerate}
\end{theorem}
{\IEEEproof} Omitted due to space limitation.
{\endIEEEproof}
\section{CONCLUSION}
\par The solution of the CRDF subject to a reconstruction kernel which is a convolution of causal kernels is presented, on abstract alphabets. Some of its properties are also presented. It is believed that the optimal reconstruction kernel as a convolution of causal kernels has several implications in applications where causality of the decoder as a function of the source is of concern. Specific example by invoking (\ref{ex13}) will be part of the final paper.


\bibliographystyle{IEEEtran}

\bibliography{photis}

\end{document}